\documentclass[aps,prl,twocolumn]{revtex4-1}
\pdfoutput=1
\usepackage[T1]{fontenc}
\usepackage[english]{babel}
\usepackage{float}
\usepackage[utf8]{inputenc}
\usepackage{pslatex}   % for times new roman font
\usepackage{amsmath}

\makeatletter

\newcommand{\Felec}{F_\mathrm{elec}}

\newcommand{\Vbias}{V_\mathrm{bias}}
\newcommand{\Vcpd}{V_\mathrm{cpd}}

\newcommand{\dcdz}{\frac{\partial C}{\partial z}}

\newcommand{\Vdc}{V_\mathrm{dc}}
\newcommand{\Vac}{V_\mathrm{ac}}

\newcommand{\df}{\Delta f}

\newcommand{\Fw}{F_\mathrm{\omega}}

\newcommand{\Fin}{F_\mathrm{in}}
\newcommand{\Fout}{F_\mathrm{quad}}
\newcommand{\wm}{\omega_\mathrm{m}}
\newcommand{\wel}{\omega_\mathrm{el}}
\newcommand{\fm}{f_\mathrm{m}}
\usepackage{babel}
\usepackage{graphicx}

\makeatother
\begin{document}

\date{\today}

\title{Kelvin probe force microscopy by direct dissipative electrostatic 
force modulation}

\author{Yoichi Miyahara}
\email[Corresponding author:]{yoichi.miyahara@mcgill.ca}
\author{Jessica Topple }
\author{Zeno Schumacher}
\author{Peter Grutter}
\affiliation{Department of Physics,
 Faculty of Science, McGill University, Montreal, Quebec, Canada H3A 2T8}

\begin{abstract}
We report a new experimental technique for Kelvin probe force
microscopy (KPFM) using the dissipation signal of frequency modulation 
atomic force microscopy for bias voltage feedback.
It features a simple implementation and faster scanning
as it requires no low frequency modulation.
The dissipation is caused by the oscillating electrostatic force
that is coherent with the tip oscillation,
which is induced by a sinusoidally oscillating voltage
applied between the tip and sample.
We analyzed the effect of the phase of the oscillating force
on the frequency shift and dissipation
and found that the relative phase of 90$^\circ$ 
that causes only the dissipation is the most
appropriate for KPFM measurements.
The present technique requires a significantly smaller ac voltage amplitude 
by virtue of enhanced force detection due to the resonance enhancement
and the use of fundamental flexural mode oscillation 
for electrostatic force detection.
This feature will be of great importance 
in the electrical characterizations of technically relevant materials
whose electrical properties are influenced 
by the externally applied electric field 
as is the case in semiconductor electronic devices.
\end{abstract}

\maketitle

\section*{Introduction}
Kelvin probe force microscopy (KPFM), a variant of atomic force microscopy
(AFM) has become one of the indispensable tools 
used to investigate electronic properties of nanoscale material as well as
nanoscale devices.
In KPFM, a surface potential of a sample is measured 
by detecting a capacitive electrostatic force that is a function of 
the surface potential and applied bias voltage.
In order to separate the electrostatic force component from other force 
components such as van der Waals force, chemical bonding force 
and magnetic force, 
the electrostatic force is modulated by applying an ac bias voltage 
and the resulting modulated component of the measured force is detected
by lock-in detection \cite{Nonnenmacher91}.

KPFM has been implemented in a variety of ways that can be classified 
into two distinct categories, amplitude modulation (AM) 
\cite{Kikukawa96, Sommerhalter1999, Zerweck05} 
and frequency modulation (FM) \cite{Kitamura1999, Zerweck05}.
The former implementation takes advantage of enhanced electrostatic
force detection sensitivity 
by tuning the modulation frequency to one of the resonance frequencies of 
the AFM cantilever, leading to an enhanced detection of the electrostatic force 
by its quality ($Q$-) factor that can reach over 10,000 in vacuum
\cite{Schumacher2015}.

The latter method (FM-KPFM) detects the modulation 
in the resonance frequency shift
that is induced by a low frequency ac voltage. 
While this method requires a much higher ac voltage amplitude, 
it offers higher spatial resolution because the resonance frequency shift is 
determined by the electrostatic force gradient 
with respect to the tip-sample distance
rather than the force itself \cite{Sommerhalter1999, Zerweck05, Burke09a}.

Here we report a new KPFM implementation (D-KPFM) 
using the dissipation signal of conventional FM-AFM systems 
to detect the electrostatic force. 
The dissipation arises from an oscillating electrostatic force 
acting on the AFM tip that is coherent with the tip oscillation (trajectory). 
The oscillating electrostatic force is induced by applying an ac voltage 
between the tip and sample. 
By setting the phase of the ac voltage to $90^\circ$ out of phase 
with respect to the tip oscillation, an electrically induced dissipation 
can result without affecting the resonance frequency shift, 
allowing topography imaging to be performed 
in constant frequency shift mode. 
The KPFM feedback loop can be implemented with the electrically induced dissipation signal 
as it is proportional to the effective dc potential difference 
between the tip and sample, $(\Vbias -\Vcpd)$,  
where $\Vbias$ and $\Vcpd$ are an applied dc voltage 
and the contact potential difference, respectively.

This new technique makes it possible to take full advantage 
of the enhanced force sensitivity 
by the high $Q$-factor fundamental resonance mode
as there is no need to excite higher flexural modes
for the electrostatic force detection.
In conventional AM-KPFM 
\cite{Kikukawa96, Sommerhalter1999},
because the fundamental mode is used for topography imaging,
the frequency of ac bias voltage is typically configured 
to excite a higher resonance mode
(typically second flexural mode)
that have much higher effective spring constant,
cancelling the high $Q$-factor resonance enhancement.
The enhanced sensitivity of the technique presented here
allows the use of a much smaller ac voltage 
enabling less invasive potential measurements,
which is of great importance 
in the electrical characterizations of technically relevant materials
whose electrical properties are influenced 
by the externally applied electric field 
as is the case in semiconductor electronic devices.

Although a similar technique has already been reported 
by Fukuma \textit{et al.} \cite{Fukuma2004},
it has not been adopted widely probably because of complexity 
in its implementation.
The present technique is simpler in implementation
and even requires no additional lock-in amplifier
as there is no low-frequency modulation involved. 

\section*{Theory}
The electrostatic force between two conductors connected 
to an ac and dc voltage source, $\Felec$,
is described as follows \cite{Kantorovich00}:
\begin{eqnarray} 
       \Felec &=& -\frac{1}{2}\dcdz \{\Vbias -\Vcpd +\Vac \cos (\wel t + \phi)\}^2 \nonumber\\ 
       & = & \alpha\{\Vdc + \Vac\cos (\wel t + \phi)\}^2 \label{eq:Felec}\nonumber\\
       & = & F_0 + F_1 + F_2 
\end{eqnarray}
\[ 
  \label{eq:alpha}
         \alpha \equiv -\frac{1}{2}\dcdz \nonumber, \,\,\,  \Vdc \equiv \Vbias - \Vcpd
\]
where $C$ is the tip-sample capacitance, 
$\Vbias$ and $\Vcpd$ are the applied dc voltage and the contact potential difference,
and $\Vac$, $\wel$ and $\phi$ are the amplitude, angular frequency and phase 
of the ac bias voltage.
$z$ is the position of the tip with respect to the sample surface 
and the oscillating tip around its mean position, 
$z_0$, expressed as $z(t)= z_0 + A\cos(\wm t)$ 
with $\wm$ and $A$ being its oscillation angular frequency and amplitude,
respectively.
We assume that the tip oscillation is driven by another means 
such as piezoacoustic or photothermal excitation
and its frequency, $\fm=\wm/2\pi$, 
is chosen to be that of the fundamental flexural resonance mode, $f_0$.

Expanding Eq.~\ref{eq:Felec} and isolating each harmonic component, 
$F_0$, $F_1$ and $F_2$, 
which result from the applied ac bias voltage, yields the following terms:
\begin{eqnarray} 
       F_0 & = & \alpha \left(\Vdc ^2 + \frac{\Vac^2}{2} \right)\label{F0} \\ 
       F_1 &= & 2\alpha \Vdc \Vac \cos (\wel t + \phi)  \label{F1} \\ 
       F_2 & = & \frac{1}{2}\alpha \Vac^2 \cos \{2(\wel t + \phi)\}  \label{F2} 
  \end{eqnarray}

Notice that the $z$-dependence of $\alpha = -\frac{1}{2}\dcdz(z)$ 
must be taken into account
in order to correctly describe the response of the oscillating cantilever 
subject to the oscillating electrostatic force induced 
by the coherent ac voltage.

By expanding $\alpha$ around the mean position, $z_0$, and taking the first order term,
$\alpha$ is expressed as follows:

 \begin{equation}
  \alpha (z) \approx \alpha(z_0) + \alpha ' (z-z_0) = \alpha_0 + \alpha ' A \cos(\wm t)
  \label{eq:alpha}
\end{equation}

Substituting eq.~\ref{eq:alpha} into eq.~\ref{F0} and rearranging it yields, 
\begin{eqnarray}
       F_0(t)  &= & \{\alpha_0 + \alpha' A \cos(\wm t)\} \left(\Vdc ^2 + \frac{\Vac^2}{2}\right) \nonumber \\
       &=& \alpha_0 \left(\Vdc ^2 + \frac{\Vac^2}{2}\right)  
       + \alpha' A\left(\Vdc ^2 + \frac{\Vac^2}{2}\right) \cos (\wm t)
      \label{Fdc2} 
\end{eqnarray}

Note that the second term expresses an oscillating force whose frequency is $\wm/2\pi$
while the first term expresses the static deflection of the cantilever.
As this oscillating force is in phase with respect to $z(t)$, 
it results in a shift in resonance frequency \cite{Holscher2001}.

Substituting eq.~\ref{eq:alpha} into the expression of $F_1$ (Eq.~\ref{F1})
gives
% Omega term
\begin{align}
       F_1(t) & =  2 \{\alpha_0 + \alpha' A \cos(\wm t)\} \Vdc\Vac \cos (\wel t + \phi) \nonumber\\
               =   2& \alpha_0  \Vdc\Vac \cos(\wel t + \phi) \nonumber\\
      +   \alpha' &A\Vdc\Vac [\cos\{(\wel + \wm) t + \phi)\}
      +  \cos\{(\wel - \wm)t -\phi\}]
       \label{eq:F1} 
\end{align}
When $\wel \ll \wm$, a pair of sideband peaks appear 
around the mechanical resonance peak 
due to the second and third terms of eq.~\ref{eq:F1}, 
which are detected in the conventional FM-KPFM
\cite{Zerweck05}.

We focus on a special case where $\wel = \wm$.
$F_1(t)$ can be simplified as follows: 
\begin{align}
       F_1(t) & =  \alpha' A\Vdc\Vac \cos \phi + 2 \alpha_0  \Vdc\Vac \cos(\wm t + \phi) \nonumber \\       +  & \alpha' A\Vdc\Vac \cos(2\wm t + \phi) 
       \label{eq:F1_1} 
\end{align}
We notice that the interaction between the mechanical oscillation of the tip and 
the oscillating electrostatic force produce a static force and 
two harmonic forces with frequency, $\wm$ and $2\wm$.

Likewise $F_2(t)$ (eq.~\ref{F2}) contains three harmonic terms 
with their frequency, $\wm$, $2\wm$ and $3\wm$ 
under the same condition ($\wm = \wel$) as follows:
\begin{eqnarray}
  \label{eq:F2}
       F_2(t) &=& \frac{1}{2} \{\alpha_0 +  \alpha' A \cos(\wm t)\} \Vac ^2 \cos \{2(\wel t + \phi)\} \nonumber \\
       & = & \frac{1}{2}  \alpha_0 \Vac ^2 \cos \{2(\wm t + \phi)\} \nonumber\\
      & + &\frac{1}{4} \alpha' A \Vac ^2 \{ \cos (\wm t + 2\phi)
       +  \cos (3\wm t + 2\phi)\} 
\end{eqnarray}

As is shown in the theory of FM-AFM 
\cite{Holscher2001,Kantorovich2004, Sader2005},
the resonance frequency shift and dissipation signal
are essentially determined by the in-phase and quadrature component of the 
fundamental harmonic component (in this case $\wm$) of the oscillating 
force, respectively.
Putting together all the $\wm$ components from $F_0$, $F_1$ and $F_2$
and rearranging them,
we get the following expression:

\begin{equation}
       \label{Fw} 
       \Fw(t)  =  \Fin \cos \wm t + \Fout \sin \wm
\end{equation}

where
\begin{eqnarray}
  \Fin & = & \alpha'A \left(\Vdc + \frac{\alpha_0 \cos \phi}{ \alpha' A} \Vac \right)^2 \\
    & & - \left[\frac{\alpha_0^2 \cos^2 \phi}{\alpha'A} 
      - \frac{\alpha'A}{2}\left(1+ \frac{\cos 2\phi}{2}\right)\right]\Vac^2  \label{eq:Fin}
\end{eqnarray}

\begin{equation}
  \Fout =  -2\alpha_0 \Vdc\Vac  \sin(\phi) - \frac{1}{2}\alpha'A\Vac^2 \sin(\phi)\cos(\phi)
  \label{eq:Fout}
\end{equation}

These $\Fin$ and $\Fout$ cause the resonance frequency shift and dissipation
signal in FM-AFM, respectively.
As can be seen in the formula for $\Fin$, 
the frequency shift versus $\Vdc$ curve will be a parabola whose minimum 
is shifted from $\Vcpd$
by the value determined by the phase, $\phi$,  and amplitude, $\Vac$, of
the oscillating bias voltage.
This indicates that the bias voltage at the parabola minimum is no longer $\Vcpd$. 
However, in the special case where $\phi = 90^\circ$,
we find 
\begin{eqnarray}
  \Fin & = &\alpha' A \left(\Vdc ^2 + \frac{\Vac^2}{4}\right)  \label{eq:Fin90}\\
  \Fout & =& -2\alpha_0 \Vdc\Vac = -2\alpha_0 (\Vbias - \Vcpd)\Vac \label{eq:Fout90}
\end{eqnarray}

The resulting resonance frequency shift, $\Delta f$ and
the dissipation signal, $g$, are obtained using the formulas found
in Ref.~\citenum{Holscher2001,Kantorovich2004, Sader2005} as follows:

\begin{align}
  \Delta f & = -\frac{1}{2}\frac{f_0}{k}\frac{\Fin}{A}
  = -\frac{1}{2}\frac{f_0}{k}\alpha' \left\{(\Vbias - \Vcpd)^2 + \frac{\Vac^2}{4}\right\}\\
  g & = g_0 \left(1 - \frac{Q}{kA}\Fout\right)
   = g_0\left\{1 + 2\frac{Q}{kA}\alpha_0(\Vbias - \Vcpd)\Vac\right\}
      \label{eq:g}
\end{align}
where $k$ is the effective spring constant 
of the fundamental flexural mode of the cantilever
and $g_0$ is the dissipation without the ac bias voltage
which is given by the mechanical $Q$ factor of the cantilever.
%(Derivation of $\df$ and $g$ is found in Supplementary information.).
In the case where the applied bias, $\Vbias$, is equal to 
the contact potential difference, $\Vcpd$, 
the dissipation goes back to its original value, $g_0$.
It is therefore possible to use the dissipation signal, $g$, 
as the KPFM bias voltage feedback signal
with $g_0$ as its control setpoint value.
%  Thus, measuring the dissipation signal, $g$,
%  makes it possible to measure $\Vcpd$ by using $g$ 
% for the bias voltage feedback.

\section*{Experimental}
\begin{figure}[t]
  \centering
  \includegraphics[width=80mm]{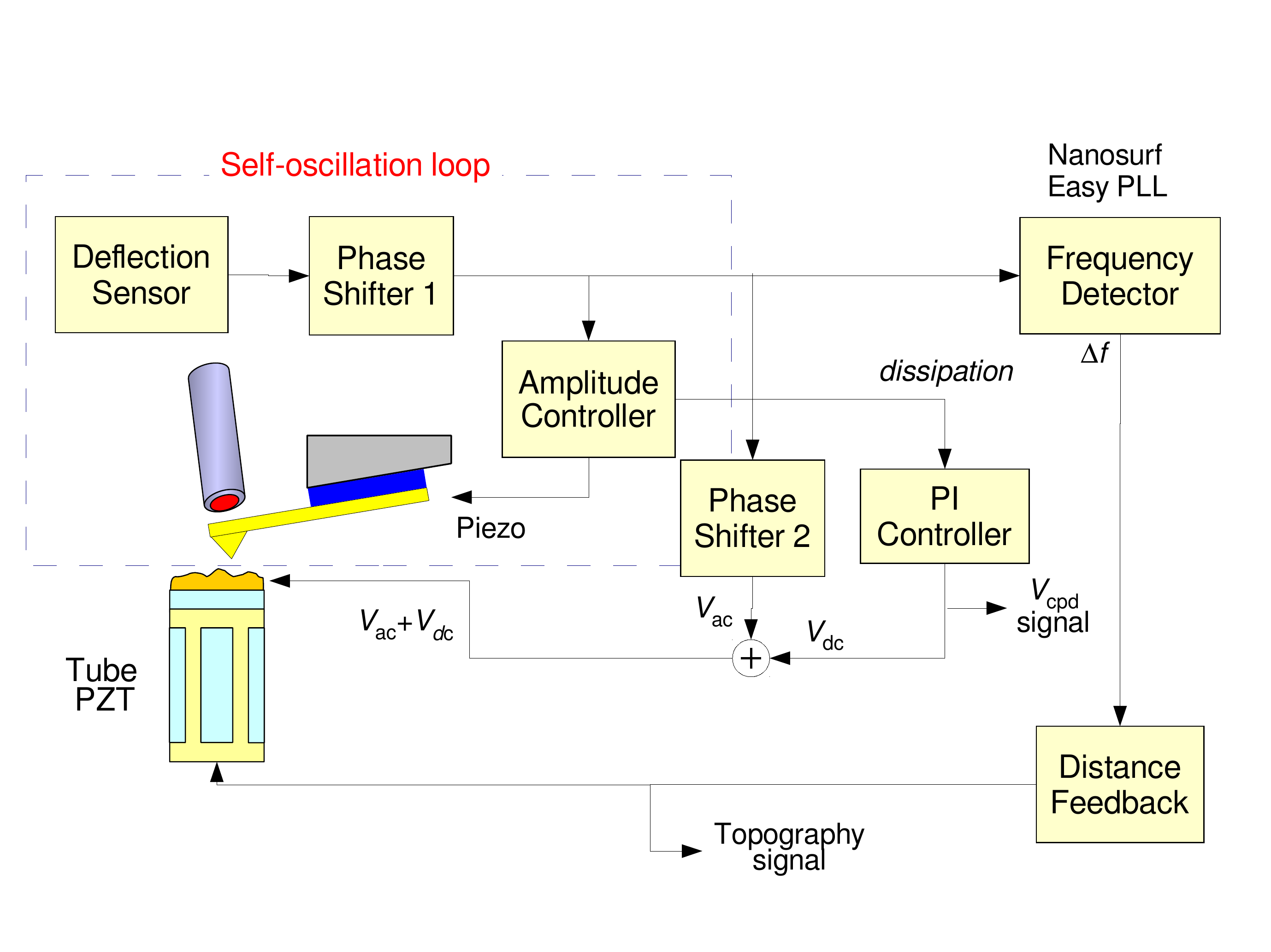}
  \caption{Block diagram of the experimental setup for D-KPFM measurements.
    \label{fig:DKPFMDiagram2}}
\end{figure}

Figure~\ref{fig:DKPFMDiagram2} depicts the block diagram of the experimental 
setup used for D-KPFM measurements.
As we notice, the D-KPFM technique requires only two additional components,
a phase shifter and proportional-integrator (PI) controller
compared to normal FM-AFM systems.
The oscillation of the AFM cantilever was controlled 
by a self-oscillation feedback loop electronics 
which consists of a phase shifter and an amplitude controller.
The amplitude controller is used to maintain a constant oscillation amplitude
and composed of a root-mean-square (RMS) amplitude detector and
a PI controller (NanoSurf easyPLLplus oscillator controller).
The output of the amplitude feedback PI controller is the dissipation signal
which will be used for controlling the dc bias voltage.
The detection bandwidth of the RMS amplitude detector 
was extended to about 10~kHz 
by replacing the integration capacitor in the original RMS detector circuit.

The deflection signal is fed into the additional phase shifter, 
which serves to adjust the relative phase, $\phi$,
to produce the ac voltage which is $90^\circ$ out of phase 
to the cantilever deflection.
Due to the phase delay in the deflection sensing electronics,
the actual phase shift value set by the phase shifter may be
different from $90^\circ$.
% Although in the following experiments 
% we used a digital lock-in amplifier 
% (HF2LI, Zurich Instruments)
% operated in phase-locked loop mode as a phase shifter for convenience,
% other simpler phase shifter circuits such an all-pass filter
% can also be used for the D-KPFM measurement.
The dissipation signal acts as the input signal to the PI controller,
which adjusts the applied dc bias, $\Vbias$, 
to maintain a constant dissipation equal to the value without $\Vac$ applied,
$g_0$.

We used a JEOL JSPM-5200 atomic force microscope for the experiments 
with the modifications described below.
The original laser diode was replaced by a fiber-optic collimator 
with a focusing lens that is connected 
to a fiber-coupled laser diode module (OZ Optics).
The laser diode was mounted on a temperature controlled fixture 
and its driving current was modulated with a radio frequency signal 
to reduce the deflection detection noise \cite{Fukuma05}.
The bias voltage was applied to the sample 
with reference to the grounded AFM tip 
to reduce the effect of the capacitive crosstalk of $\Vac$ 
to the cantilever excitation piezo \cite{Melin2011,Diesinger2012}.
The excitation piezo was shunted with a chip resistor 
with low resistance ($\sim 10$~$\Omega$) to further reduce the effect.
The original controller was replaced with a open-source controller, 
GXSM, \cite{Zahl2010}
with the dedicated acquisition hardware (MK2-A810, SoftdB).

A commercial silicon AFM cantilever (NSC15, MikroMasch) 
with a typical spring constant of 20~N/m and resonance frequency 
of $\sim 300$~kHz was used in high-vacuum environment with
the pressure of $1 \times 10^{-7}$~mbar.
The oscillation amplitude and quality factor of the cantilever 
used for the D-KPFM imaging were 5~nm$_\text{p-p}$ and $\sim 5500$, 
respectively.

\section*{Results and Discussion}
\begin{figure}[t]
  \centering
  \includegraphics[width=80mm]{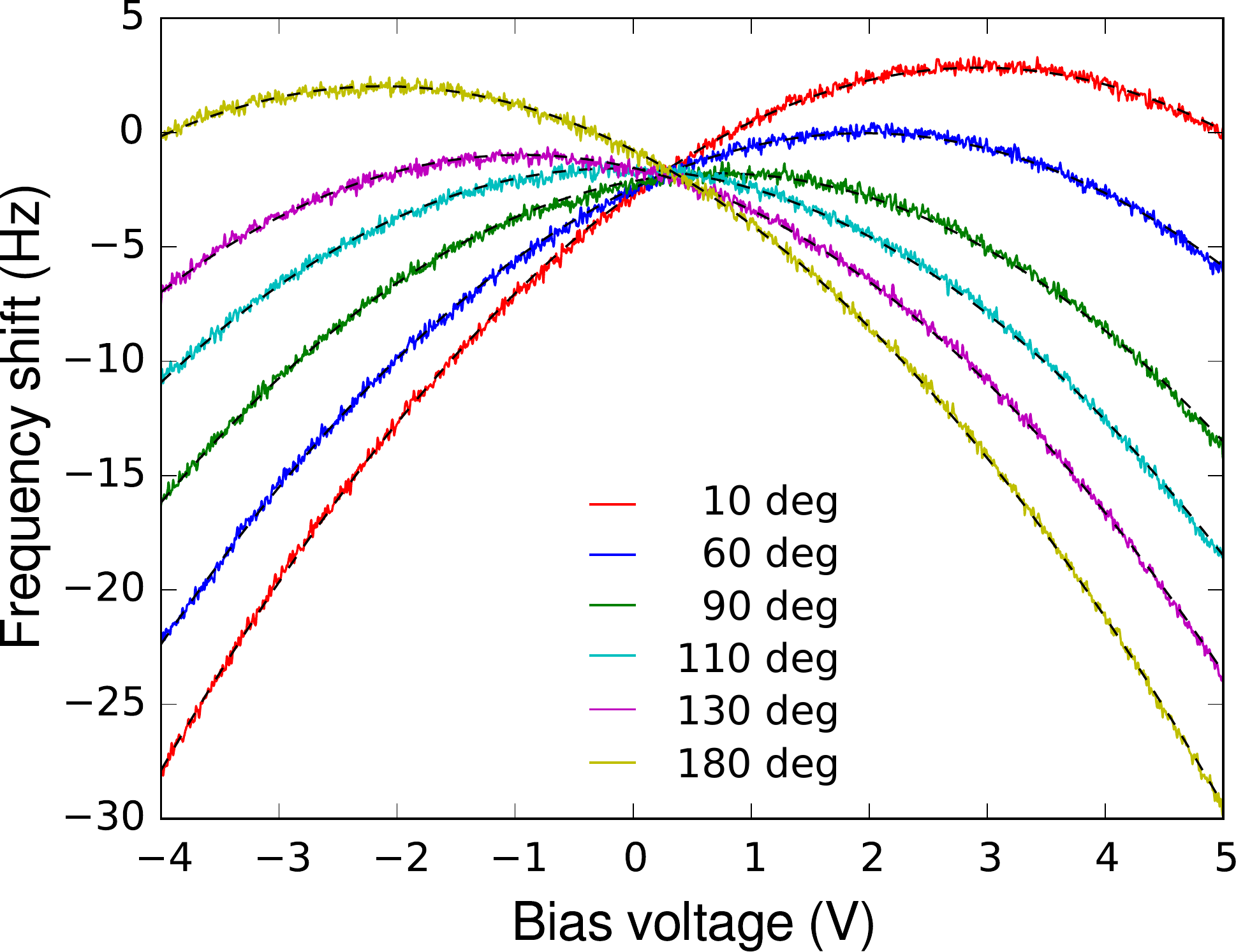}\\
  \includegraphics[width=80mm]{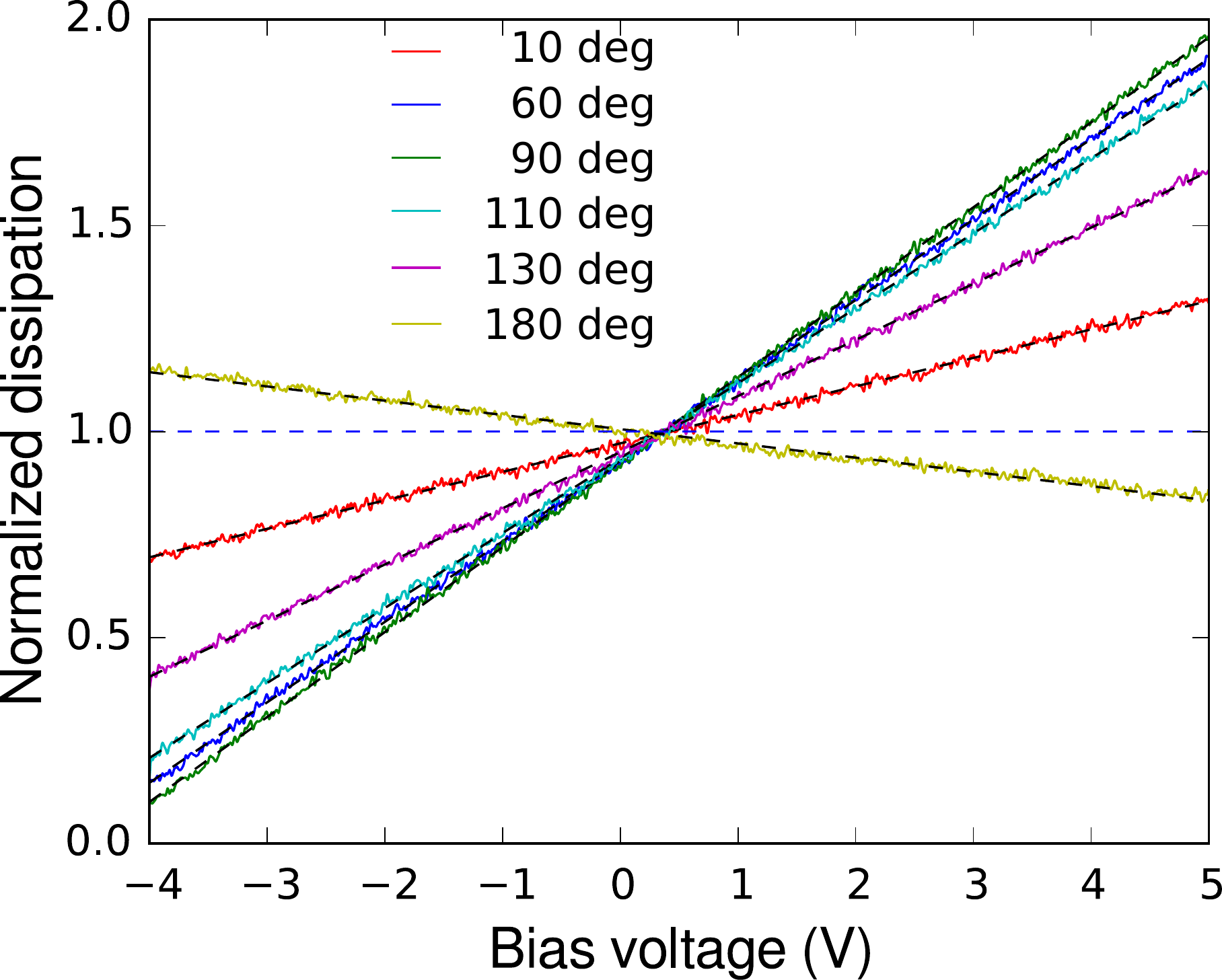}   
  \caption{(a) Frequency shift, $\df$, and (b) dissipation signal, $g$, 
  versus dc bias voltage, $\Vbias$, curves 
  taken with a coherent sinusoidally oscillating voltage 
  with the amplitude, $\Vac=100$~mV$_\text{p-p}$ and various phases, $\phi$,
  applied to a 200~nm thick SiO$_2$ on Si substrate.
  The dissipation signal is normalized with the value without 
  the ac bias voltage (indicated with the horizontal blue dashed line).
  In both figures, each of dashed lines represent fitted curves 
  assuming a parabola for $\df$ and a linear line for $g$ as indicated
  in eq.~\ref{eq:Fin} and \ref{eq:Fout}, respectively.
  The oscillation amplitude of the tip was 7.2~nm$_\text{p-p}$ 
  and the quality factor of the cantilever was 9,046.
  \label{fig:df-V}}
\end{figure}

Figure~\ref{fig:df-V} shows simultaneously measured 
$\df$ and $g$ versus $\Vbias$ curves 
with a coherent sinusoidally oscillating voltage 
with the amplitude, $\Vac=100$~mV$_\text{p-p}$ and various phase, $\phi$.
The curves were taken on a Si substrate with natural oxide SiO$_2$.
A fitted curve with a parabola for $\df$-$\Vbias$ curves
(eq.~\ref{eq:Fin})
or a linear line for $g$-$\Vbias$ curves (eq.~\ref{eq:Fout}) 
is overlaid on each experimental curve,
indicating a very good agreement between the theory and experiments.
As can be seen in Fig.~\ref{fig:df-V}(a) and (b), 
the position of the parabola vertex shifts 
and the slope of $g$-$\Vbias$ curve changes 
systematically with varying phase. 

\begin{figure}[tt]
  \centering
  \includegraphics[width=80mm]{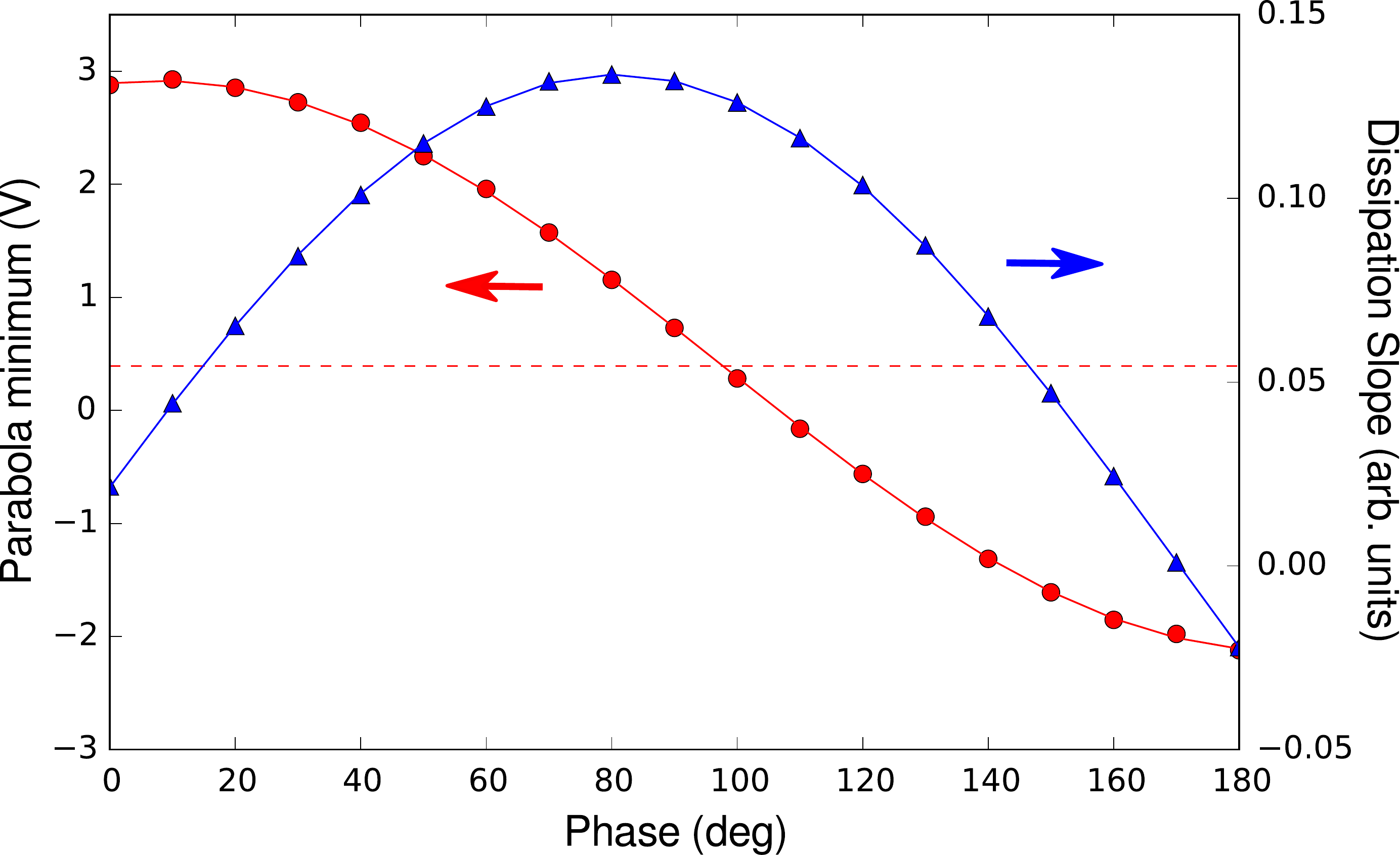}   
  \caption{Voltage of minimum of the measured $\df$-$\Vbias$ curves 
  (red circles) (Fig.~\ref{fig:df-V}(a))
  and the slope of dissipation-$\Vbias$ curves (blue circles)
  (Fig.~\ref{fig:df-V}(b)). 
  Each solid line represents the fitted curve with the cosine function (eq.~\ref{eq:Fin})
  for the parabola minimum and with the sine function (eq.~\ref{eq:Fout})
  for the dissipation slope.
  The horizontal dashed line indicates the voltage for parabola minimum without 
  the ac bias voltage.
  \label{fig:fit_parameters}}
\end{figure}

In order to further validate the theoretical analysis, 
the voltage for parabola maximum of $\df$-$\Vbias$ curves and 
the slope of $g$-$\Vbias$ curves are plotted against the phase, $\phi$,
in Fig.~\ref{fig:fit_parameters}.
Each plot is overlaid with a fitted curve (solid curve) with the cosine function
%Each overlaid solid line is the fitted curve with the cosine function 
(eq.~\ref{eq:Fin})
for the parabola maximum and with the sine function (eq.~\ref{eq:Fout}) 
for the dissipation slope,
demonstrating an excellent agreement between the experiment and theory.
The voltage for parabola maximum versus phase curve intersects 
the value for the parabola without ac bias voltage at the phase of 97$^\circ$ 
as opposed to 90$^\circ$ which is predicted by the theory.
This deviation is mainly due to the phase delay 
in the photo-diode preamplifer electronics.
The dissipation slope takes its maximum value at around 81$^\circ$, 
again deviating from the theoretical value of 90$^\circ$.
%which is supposed to be 90$^\circ$ theoretically.
This deviation is probably due to the residual capacitive crosstalk 
to the excitation piezo \cite{Melin2011,Diesinger2012}.

\begin{figure}[t]
  \centering
  \includegraphics[width=88mm]{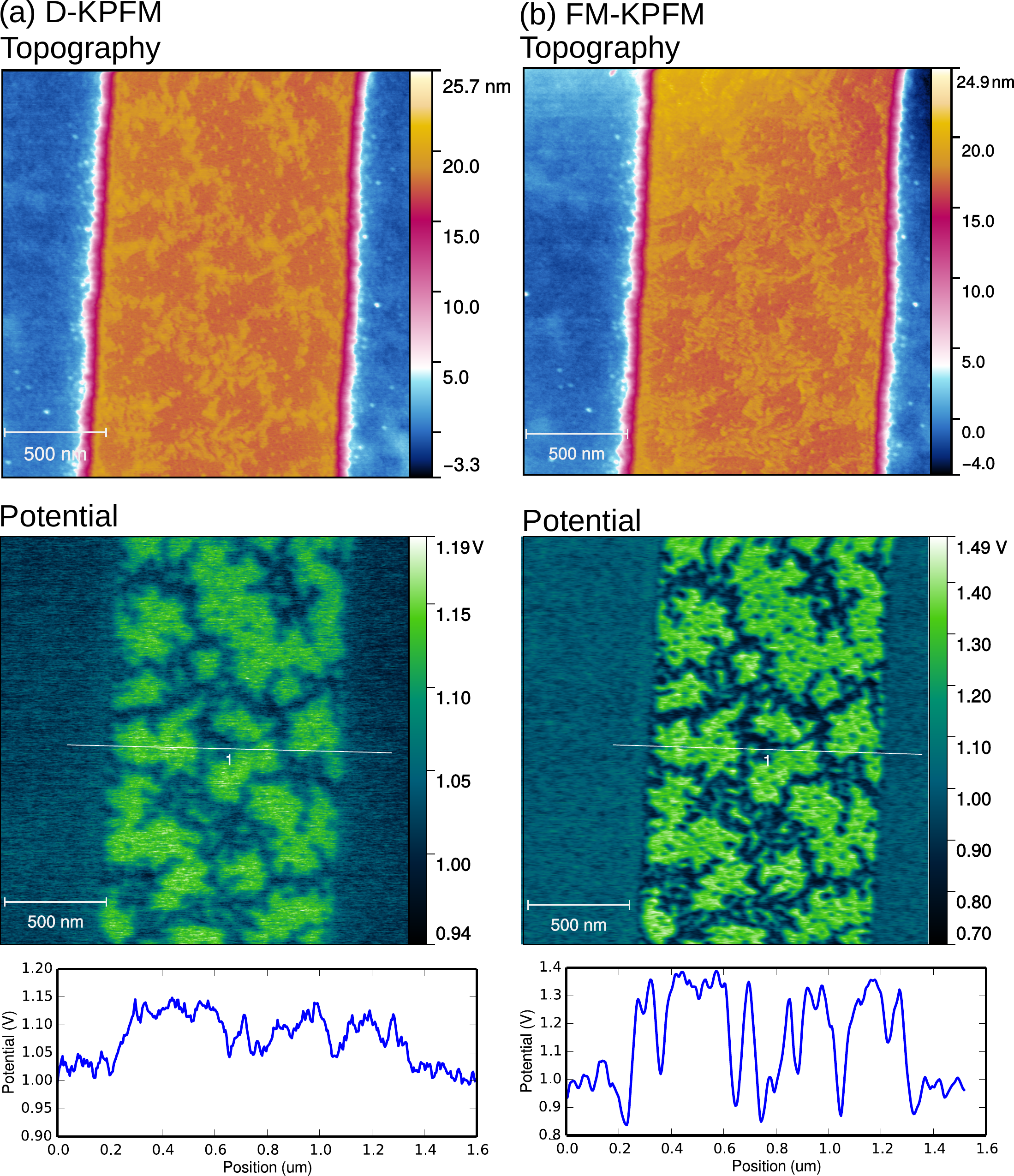}
  \caption{Simultaneously taken topography and potential images of patterned MoS$_2$ on SiO$_2$/Si substrate 
        by (a) D-KPFM ($\df = -4.2$~Hz, $A=5$~nm$_\text{p-p}$, $\Vac=80$~mV$_\text{p-p}$) and 
        (b) FM-KPFM techniques ($\df = -4.2$~Hz, $A=5$~nm$_\text{p-p}$, 
        $\Vac=1.0$~V$_\text{p-p}$, $f_\text{ac}=300$~Hz).  \label{fig:DKPFM_image}}
\end{figure}

Figure~\ref{fig:DKPFM_image} shows topography and potential images 
of a patterned MoS$_2$ on SiO$_2$/Si substrate taken 
by (a) D-KPFM and (b) FM-KPFM techniques with the same tip.
In D-KPFM imaging, a sinusoidally oscillating voltage 
with an amplitude of $\Vac=80$~mV$_\text{p-p}$ 
phase-locked with the tip oscillation was applied to the sample.
In FM-KPFM imaging, a sinusoidally oscillating voltage 
with the amplitude of $\Vac=1.0$~V$\text{p-p}$
and frequency of 300~Hz was applied to the sample.
The scanning time for D-KPFM and FM-KPFM imaging were 
1 and 2~s/line, respectively.
The number of pixels of the images is 512$\times$512.

A flake of MoS$_2$ was deposited onto a SiO$_2$/Si substrate 
by mechanical exfoliation
and a stripe pattern was created by reactive ion etching on top of the flake.
The topography images show an unetched ridge located 
between the etched regions. 
The height of the ridge is approximately 20~nm 
with respect to the etched regions.
A clear fractal-like pattern can be seen on the ridge in both potential images.
The potential contrast can be ascribed to the residue of the etch resist (PMMA)
as the topography images show the similar contrast whose thickness is about 1~nm.

Although both potential images taken with D-KPFM and FM-KPFM show 
a very similar pattern on the ridge,
we notice lower contrast in the potential image taken 
with D-KPFM than that with FM-KPFM.
The difference is more clearly seen in the line profile 
attached for each potential image.
The peak-to-peak value of the potential variation in the D-KPFM image 
is $\sim 0.15$~V,
about one third that in the FM-KPFM image ($\sim 0.5$~V).
The similar difference has been observed 
in the potential contrast taken with FM-KPFM and AM-KPFM
and is ascribed to the fact that
the AM-KPFM is sensitive to electrostatic force 
whereas FM-KPFM uses the modulation in the resonance frequency shift 
which is sensitive to force gradient 
\cite{ Zerweck05, Glatzel2003a}.
The similarity between D-KPFM and AM-KPFM is apparent
%in $\Fout$ (Eq.~\ref{eq:Fout})
in the expression of $g$ (eq.~\ref{eq:g})
which is proportional 
to $\alpha_0 = -\frac{1}{2}\frac{\partial C}{\partial z}|_{z_0}$
rather than $\alpha'$.
This indicates that the smaller potential variation observed in D-KPFM 
resulted from larger spatial average due to the stray capacitance
including the body of the tip and the cantilever \cite{Hochwitz1996,Jacobs98}.

In spite of lower contrast, D-KPFM has a clear advantage 
that it requires much smaller $\Vac= 80~$mV$_\text{p-p}$ 
compared with 1~V$_\text{p-p}$ for FM-KPFM, in this case.
This advantage is important for such samples as semiconductor 
where the influence of the large $\Vac$ can be very important 
due to band-bending effects.

The detection bandwidth of D-KPFM is determined 
by the bandwidth of the amplitude control feedback loop used in FM-KPFM.
In fact, applying the coherent $\Vac$ causing dissipative force can be used 
to measure the dynamics of the amplitude control feedback system.
In FM-KPFM, 
the AFM cantilever serves for the frequency determining element 
of an oscillator circuit (i.e.self-oscillator) 
so that the oscillation frequency of the oscillator keeps track 
of the resonance frequency of the cantilever.
In this way, the conservative force has no influence on the drive amplitude
\cite{Labuda2011}
and the amplitude controller compensates for 
the effective $Q$ factor change caused by dissipative force.
Therefore, by modulating $\Vbias$ at a low frequency (< a few kHz)
together with applying the coherent $\Vac$, 
the amplitude of the dissipative force can be modulated 
as can been seen in eq.~\ref{eq:Fout90}
and the frequency response of the amplitude feedback loop can thus be measured
with a lock-in amplifier.
The measured $-3$~dB bandwidth of the amplitude feedback loop 
is as high as 1~kHz,
which is wider than that of the PLL frequency detector (400~Hz).
Note that in contrast to the settling time of the oscillation amplitude
of a cantilever subject to a change in conservative force
which is $\tau \sim Q/f_0$ \cite{Albrecht1991},
the response time to a dissipative force is not limited by $Q$ 
and can be faster as the energy dissipated per cycle is given by $\pi \Fout A$
rather than $\pi k A^2/Q$ for the case of conservative force.
This explains the observed fast response of the amplitude feedback loop,
resulting in the wider bandwidth of the voltage feedback loop in D-KPFM
than that in FM-KPFM that is limited by PLL demodulation bandwidth
(typically $ <1$~kHz) which sets the bias modulation frequency.

The noise of $\Vcpd$ is ultimately determined by 
the noise in the tip oscillation amplitude, $\delta A$.
The change in the oscillation amplitude of the self-excited cantilever, 
$\Delta A$, caused by the dissipative force, $\Fout$, 
is given by
\begin{equation}
  \label{eq:Amp_dep}
 \Delta A = -\frac{Q}{k}\Fout = -\frac{Q}{k}\alpha_0\Vdc\Vac
= \frac{Q}{k}\left . \dcdz \right |_{z_0}\Vdc\Vac
  \end{equation}

Eq.~\ref{eq:Amp_dep} resembles  
the amplitude response of a simple harmonic oscillator driven on resonance
on which AM-KPFM is based.
%(See Supplementary information for the derivation).

The noise in $\Vcpd$, $\delta \Vcpd$ can thus be expressed as follows:
\begin{equation*}
  \delta \Vdc = \frac{k}{Q}\frac{\delta A}{\left .\dcdz \right|_{z_0} \Vac}
\end{equation*}
which agrees with the result by Fukuma \textit{et al.} \cite{Fukuma2004}.
This indicates that $\delta \Vcpd$ is proportional to $k/Q$.
In typical AM-KPFM measurements \cite{Glatzel2003a} 
where the second flexural mode oscillation is used 
for detecting electrostatic force,
the improvement of $\delta \Vcpd$ by the enhanced $Q$ factor 
is partially cancelled 
by the substantially higher dynamic spring constant of the second mode 
with $k_\text{2nd} \approx 40 k$ \cite{Melcher07}. 
D-KPFM enables to fully take advantage of the resonance enhancement 
while retaining the advantages of the single-pass FM-AFM.

So far we assumed that no other process other
than the dissipative electrostatic force causes the dissipation signal
in FM-AFM systems. 
Although it is often the case 
that the contribution from other processes are negligible,
even when other intrinsic or extrinsic dissipation processes are present,
it is possible to separate the dissipation induced by $\Fout$ 
just as is done in FM-KPFM with $\Delta f$ 
or in the method by Fukuma \textit{et al.} \cite{Fukuma2004}.
In this case $\Vbias$ needs to be modulated at a low frequency 
and the resulting modulated dissipation signal 
is used for the dc bias feedback.
Clearly this scheme is slower than the fast response achieved 
by D-KPFM technique.

In conclusion, we report a new experimental technique for Kelvin probe force
microscopy using the dissipation signal of FM-AFM for dc voltage feedback.
It features the simpler implementation and faster scanning
as it requires no low frequency modulation.
The dissipation is caused by the oscillating electrostatic force
that is coherent with the tip oscillation,
which is induced by a sinusoidally oscillating ac voltage
applied between the tip and sample.
We analyzed the effect of the phase of the oscillating force
on the frequency shift and dissipation
and found that the relative phase of 90$^\circ$ is the most
appropriate for KPFM measurements.
D-KPFM technique requires a significantly smaller ac voltage amplitude 
(a few tens of mV)
by virtue of the resonance enhanced force detection 
and the use of fundamental flexural mode oscillation.
This feature will be useful in the electrical characterizations of materials
whose electrical properties are sensitive to the externally applied electric field.

\begin{acknowledgments}
The authors would like to thank Dr.~Omid Salehzadeh Einabad 
and Prof.~Zetian Mi at McGill University for 
providing the MoS$_2$ sample.
This work was partly supported by the Natural Science and Engineering
Research Council (NSERC), 
le Fonds Qu\'eb\'ecois de Recherche sur la Nature et
les Technologies (FQRNT).
\end{acknowledgments}

\end{document}